\documentclass[11pt,a4paper]{article}
\usepackage{geometry}
\usepackage[british]{babel}
\usepackage[T1]{fontenc}
\usepackage[utf8]{inputenc}
\usepackage[pdftex]{graphicx} 
\usepackage{url}
\usepackage{fourier,charter}
\usepackage{varioref}\labelformat{equation}{(#1)}
\usepackage[round]{natbib}
\usepackage{enumerate,mathrsfs}
\usepackage{amsfonts}
\newcommand{\anymode}[1]{\ifmmode{#1}\else\mbox{$#1$}\fi}
\newcommand{\biblioitem}[1]{\par \frenchspacing
           \vbox{\noindent \hangindent=2em \hangafter=1{#1}}
           \vskip\parskip}
\def\biblioitem{\par\vskip\parskip\noindent\hangindent=2em \hangafter=1}

\newcommand\NB[1]{\kern-0.4em\raisebox{-1.5ex}{$\stackrel{\big|}{\hbox{%
  \tiny\sc NB}}$}\marginpar{\sl\footnotesize#1\hfill}\kern-0.2em}

\long\def\Nota#1{\footnote{#1}\kern-0.2em\NB{cfr Nota N.\,\arabic{footnote}}}

\def\linea#1{\ifhmode\hfill\break\fi\hbox to \hsize{#1}}

\newcommand{\inv}{^{-1}}

\def\vc{v\kern -0.1em .c.\relax}

\let\corr=\cor

\newcommand{\dfrac}[2]{\displaystyle{\frac{#1}{#2}}}

\newcommand{\E}[2][]{
   \ensuremath{\mathbb{E}_{#1}\!\left\{\displaystyle{#2}\right\}}}
\newcommand{\eps}{\varepsilon}

\long\def\ignore#1{}

\newcommand{\indep}{\perp\kern-0.5em\perp}

\newcommand{\pr}[2][]{
   \ensuremath{\mathbb{P}_{#1}\!\left\{\displaystyle{#2}\right\}}}

\newcommand{\Real}{\mathbb{R}}

\newcommand{\T}{^{\top}}
\newcommand{\var}[2][]{
   \ensuremath{\textrm{var}_{#1}\!\left\{\displaystyle{#2}\right\}}}

\newcommand{\Expn}{\mathrm{Expn}}

\newcommand{\Poisson}{\mathrm{Poisson}}

\newcommand{\N}{\mathrm{N}{}}

\renewcommand{\d}{\,\mathrm{d}}

\let\phi=\varphi
\usepackage{color}

\title{\bf Sample selection models for discrete and other non-Gaussian
   response variables}
\author{{\Large Adelchi Azzalini}\large  \\
  Department of Statistical Sciences \\
  University of Padua\\ Italy
  \and
  {\Large Hyoung-Moon Kim}\large\\
  Department of Applied Statistics\\
  Konkuk University\\
  Seoul, Korea
  \and
  {\Large Hea-Jung Kim}\large\\
  Department of Statistics\\
  Dongguk University\\
  Seoul, Korea
 }
\date{\small\today}

\begin{document}
\maketitle

\begin{abstract}
  Consider observation of a phenomenon of interest subject to selective
  sampling due to a censoring mechanism regulated by some other variable.  In
  this context, an extensive literature exists linked to the so-called Heckman
  selection model. A great deal of this work has been developed under Gaussian
  assumption of the underlying probability distributions; considerably less
  work has dealt with other distributions.  We examine a  general
  construction which encompasses a variety of distributions and allows various
  options of the selection mechanism, focusing especially on the case
  of discrete response.  Inferential methods based on the pertaining
  likelihood function are developed.
\end{abstract}
\clearpage
\section{Sample selection}

\subsection{Nature of the problem}
In observational studies, as opposed to experimental studies, a recurrent
problem is the presence, at least potentially, of a sample selection
mechanism, leading to a non-random sample from the target population.
Although in principle the term `sample selection' applies more generally,
it is commonly referred to situations where the target of a study is
the relationship between a response variable and a set of covariates,
but individuals are observed only conditionally on the outcome of a certain
selection factor, which is not independent from the variable of interest.
Such dependence between the response variable and the selection factor
generates a difference between the intended and the actual sampling
distribution, hence an inherent bias in the inferential process.

A concrete example of this situation is discussed in the pioneering work of
Heckman (1976, 1979) on the sample selection problem.  In a study on the
determinants of wages for female work, a linear regression model is introduced
in which the wage of a worker is connected to a set of determinants, such as
age, level of education, and so on.  In this situation, a selection mechanism
takes place because a fraction of the workers do not undertake a job whose
wage is below a certain threshold; this minimal wage level, called the
reservation wage, is for not fixed for all workers, but varies from subject to
subject.  Hence, for these subjects, we only observe the determinants of the
wage, without a wage value.  Clearly, plain exclusion of these cases from the
analysis would lead to a bias in the coefficients of the fitted regression
model, because the unobserved wages can be expected to be towards the lower
end of the wage range.

The sample selection problem is widespread in all areas where observational
studies are commonly in use. Social sciences in the broad sense, hence
including economics, represent historically the main domain of relevance of
the problem.  It is then not surprising that the main body of the pertaining
literature has been developed within econometrics and quantitative sociology.
Notice, however, that other research domains are not excluded. For instance,
the motivating example of the account of this theme by Copas \& Li (1997)
refers to a study of a new medical treatment where the allocation to the
standard or the new type of treatment was affected by some variable not
independent from the probability of success.

\subsection{Heckman model}  \label{s:heckman}

As already mentioned, fundamental work on the sample selection problem has
been done by Heckman (1976, 1979), of which we now summarize the key
ingredients. We phrased the exposition in a slightly different form with
respect to the original, although equivalent to it, to facilitate the
subsequent introduction of our construction.

Consider the case where the objective of interest is the study of the linear
relationship between a response variable $Y$ and a set of covariates $x$, but
there is the complication that the actual observation of $Y$ is possible when
an unobserved variable $U$ exceeds a certain threshold and the distribution of
$U$ is affected by another set of covariates $w$. Under assumption of joint
normality of $(Y,U)$ and linearity of the dependence of the mean values on the
covariates, the probability distribution associated to the $i$th subject
($i=1,\dots,n$) randomly drawn from the population is of the form
\begin{equation}  \label{eq:(Y,U)}
  \pmatrix{Y_i \cr U_i} \sim  \N_2\left(\pmatrix{ \mu_i \cr \tau_i },
               \pmatrix{\sigma^2 & \rho\sigma \cr \rho\sigma & 1}\right),
               \qquad\quad \mu_i=x_i\T\beta\,, \quad \tau_i=w_i\T\gamma \,,
\end{equation}
but observation of $Y_i$ only occurs under the condition $U_i\ge 0$;
the $x_i$ vector is $p$-dimensional and  $w_i$  is $q$-dimensional.
While $U_i$ is unobservable, what we can observe is the binary variable
\begin{equation}
   D_i=\cases{1 & if $U_i \ge0 $,\cr 0 & otherwise,}
   \label{eq:D}
\end{equation}
so that, equivalently, observation of $Y_i$ occurs only for cases with $D_i=1$.

The overall available information is therefore constituted by the set
of $d_i$ binary values, the triples $(y_i, x_i, w_i)$ for the subjects
with $d_i=1$ and by the pairs $(x_i, w_i)$ for those with $d_i=0$,
having denoted by $y_i$ and $d_i$ the actual values taken on by $
Y_i$ and $D_i$.  To compute the  implied likelihood function,
the ingredients are: (i) the probability of observing $Y_i$, namely
\begin{equation}
   \pr{D_i=1} = \Phi(\tau_i)
   \label{eq:Phi(tau)}
\end{equation}
where $\Phi$ denotes the $\N(0,1)$ distribution function, and
(ii) the probability density function of the observed $Y_i$, conditionally
on the event $D_i=1$, which after some algebraic work turns out to be
\begin{equation} \label{eq:ESN-pdf}
  f(y|D_i=1)
  =  \frac{1}{\Phi(\tau_i)\:\sigma} \:\phi(z)\:
     \Phi\left(\frac{\tau_i+ \rho z}{\sqrt{1-\rho^2}}\right),
     \qquad\quad z = \frac{y-\mu_i}{\sigma},
\end{equation}
where $\phi=\Phi'$.
Strictly speaking, we should write $f_i(y|D_i=1)$ in place of $f(y|D_i=1)$
to mark its dependence on ingredients varying with the index $i$, but this
notation would have become cumbersome if it was carried on similarly
with other terms to be introduced later. The log-likelihood function is then
\begin{equation}  \label{eq:logL-normal}
 \log L = \sum_{d_i=1} \log\left\{\Phi(\tau_i) \times f(y_i|D_i=1)\right\}
          + \sum_{d_i=0}\log\left\{1- \Phi(\tau_i)\right\}\,.
\end{equation}

To estimate the regression parameters $\beta$ appearing in \ref{eq:(Y,U)}, the
method proposed by Heckman (1976) is not directly based on this likelihood
function, although an expression leading to \ref{eq:logL-normal} is given in
his paper. In light of the limited computational resources of those years, a
simpler method is presented instead, by introducing a
correction factor in the regression model based on the expected value of
$Y_i$ conditionally of $U_i\ge0$, that is, the expected value of distribution
\ref{eq:ESN-pdf}.  After obtaining estimates of the required terms by a
probit model, a second-stage least-squares estimation is then employed on the
adjusted regression model; see also Heckman (1979).
However, this operational simplification is not
crucial; what matters more is the probability structure of the formulation.

For later reference, notice that the above formulation is built on two
stochastic ingredients.  We can take them to be $(Y_i, U_i)$ or, equivalently,
the $0$-mean `error terms' $(\eps_i, \zeta_i)$, where $\eps_i=Y_i-\mu_i$ and
$\zeta_i= U_i-\tau_i$, or even one of these error terms
and the residual of the linear projection of the other one on the first one.
Which form we consider is a matter of convenience.

Another point to annotate is that the density function \ref{eq:ESN-pdf} is of
the type denoted `extended skew-normal' in a stream of literature often
identified by the phrase `skew-symmetric distributions' or similarly
`symmetry-modulated distributions'.  A recent account of this theme is given
by Azzalini \& Capitanio (2014). See specifically Section~2.2 for a
comprehensive treatment of the extended skew-normal distribution, including
the missing algebraic details leading to \ref{eq:ESN-pdf}.  We shall make use
of the connection with that literature to introduce our formulation later on.
The connection with the skew-normal distribution has been noted by
Copas \& li (1997), although they restrict it only to the case with $\tau_i=0$.

\subsection{Non-Gaussian response variables} \label{s:non-gaussian}

The original Heckman construction is firmly linked to the assumption of joint
normality of the $(Y,U)$ variables. In practical work, this assumption is
often made even when it is unlikely to be appropriate, but there are cases
where it would be completely untenable, at least with respect to the
observable component, $Y$.
We recall briefly a few directions of work stemming from the original
Heckman construction.

An early extension of Heckman model to binary response variables has been
presented by Van de~Ven and Van Praag (1981). Their probability framework is
similar to the normal case, but instead of $Y_i$ we only observe its
dichotomized version $Y_i^*$, defined similarly to \ref{eq:D}, with $Y_i$
replacing $U_i$.  Hence $\pr{Y_i^*=1}=\Phi(\mu_i)$, analogously to
\ref{eq:Phi(tau)}. We shall return to this formulation later on.

A qualitatively different route is adopted by Terza (1998); see also Greene
(2012, Section 19.5.4). While the selection mechanism is still like before,
the observation $Y_i$ is not a function of $\mu_i$ and the error term
$\eps_i\sim\N(0,\sigma^2)$ only, like in \ref{eq:(Y,U)}, but these two
ingredients determine the parameter of a distribution from which $Y_i$ is
sampled.  For instance, if $Y_i$ is taken to be of Poisson type, we could
assume that $Y_i\sim \Poisson(\exp(\mu_i+\eps_i))$.  Hence we are now considering three
separate sources of variability, namely $(\eps_i, \zeta_i, Y_i)$.  One
implication of such a scheme is that the expression of the log-likelihood
function involves an additional integration over the distribution of $\eps_i$;
see equation (3) of Terza (1998) or (19-30) of Greene (2012).
Since this integration is typically not as friendly as those implicit in
\ref{eq:Phi(tau)} and \ref{eq:ESN-pdf}, it must be carried out numerically.

For the case of continuous response variables, a frequent criticism to
Heckman's proposal is its widely recognized sensitivity to the assumption of
normality.  To neutralize or at least to mitigate this problem,
Marchenko and Genton (2012) replace the normality assumption for $(Y_i, U_i)$
in \ref{eq:(Y,U)} by the one of a bivariate Student's $t$ distribution,
hence allowing for regulation of the
distribution tails via the degrees of freedom.  By exploiting the
above-mentioned connection with results on symmetry-modulated distributions,
the density in \ref{eq:ESN-pdf} is replaced by an `extended skew-$t$
distribution'; the factor in \ref{eq:Phi(tau)} is easy to adjust.
While this construction is of adaptive type and consequently less sensitive
to departure from normality than the original one of Heckman, it does not meet
the formal criteria of  classical robustness theory; a formulation in this
framework has been developed by Zhelonkin, Genton and Ronchetti (2016).

For the analysis of count responses, Marra and Wyszynski (2016) have recently
proposed a construction based on a copula function linking the response
variable and the  latent variable regulating selection,
$Y$ and $U$ in our notation. The construction allows a wide choice of the
copula function and of the marginal distribution of the response.
In this sense, its generality is comparable with the formulation to be
described in the rest of the present paper. However, it seems to us the use
of a copula formulation is less natural for a discrete data context
and the interpretation of the copula dependence parameter less simple
compared to the continuous context, as also noted by the authors.

Within this context, the aim of the present note the introduction of
a general formulation to extend  Heckman's original construction in
various directions. The selection mechanism which is inherent to this
situation has a natural connection with the literature on symmetry-modulated
distributions, as already recalled. This connection is more on the conceptual
than on the operational side, since here we move away from the requirement of
symmetry on the underlying, unselected distribution of the variable of
interest, which is typical of literature on symmetry-modulated distributions.
However, the re-formulation of Heckman model within the conceptual framework
of that literature facilitates the construction of a wider scheme, which we
develop in the next section, first in general terms and then in some
specific instances.

\section{A broad scheme for modelling sample selection}
\subsection{Selective sampling as a mechanism of distribution modulation}

The construction of Section~\ref{s:heckman} involves a bivariate random
variable appearing in equation \ref{eq:(Y,U)}; denote it as $(Y,U)$ without
subscripts  for notational simplicity.
An equivalent stochastic representation can be obtained via the introduction
of a random variable, $T$ say, independent  from the variable of interest.
To be specific, if we denote
\[
  Z=(Y-\mu)/\sigma , \quad \alpha=\rho(1-\rho^2)^{-1/2},
\]
then
\begin{equation} \label{eq:defineT-Normal}
  T=\alpha\,Z - (1+\alpha^2)^{1/2}\,(U-\tau)  \sim\N(0,1) \,.
\end{equation}
The pair $(Y,T)$ is algebraically equivalent to  $(Y,U)$
with the convenient feature that  $\corr{T, Y}=\corr{T, Z}=0$.
Note that $(Y,T)$  is formed via the projection of the error
term $\zeta$ on $\eps$, which is one of the equivalent ways
of expressing the  underlying stochastic terms indicated
towards the end of Section~\ref{s:heckman}.
Elementary algebra shows that the event $D=1$ in \ref{eq:D}
is equivalent to
\begin{equation} \label{eq:T<h(Z)}
   T \le\alpha Z + \tau\,(1+\alpha^2)^{1/2}.
\end{equation}
Hence the density of an observed $y$ value of $Y$, conditionally
on $D=1$, is
\begin{equation} \label{eq:ESN-pdf-new}
  f(y|D=1)
   = \frac{1}{\Phi(\tau)}\:\left[ \frac{1}{\sigma}\:\phi(z)\:
     \Phi\left(\tau\,(1+\alpha^2)^{1/2} + \alpha\, z \right)\right],
     \qquad\quad z = \frac{y-\mu}{\sigma},
\end{equation}
where  the term inside the square brackets is the product of the marginal
density of $Y$ times the  probability that $D=1$ conditionally on
$Y=y$ or, equivalently, on $Z=z$.  The denominator of the leading fraction
is the  appropriate normalizing constant because it still holds that
unconditionally $\pr{D=1}=\Phi(\tau)$; equivalently, the same fact
can be show by direct integration of the term in square brackets.
It is immediate that \ref{eq:ESN-pdf-new} coincides with \ref{eq:ESN-pdf}.

If we set $f$ to be the $\N(0,\sigma^2)$ density, $G_0=\Phi$ and
$h(y)=\tau\,(1+\alpha^2)^{1/2}+\alpha(y-\mu)/\sigma$, density
\ref{eq:ESN-pdf-new} can be re-written as an instance of the more
general form
\begin{equation} \label{eq:f0-perturbate}
      f(y|D=1) = \frac{1}{\pi} \: f(y)\: G_0\{h(y)\}
\end{equation}
with normalizing constant
 \begin{equation} \label{eq:pi}
   \pi = \int_{\Real} f(y)\: G_0\{h(y)\}\d{y}\,.
\end{equation}

In what follows, we shall consider alternative distributions of type
\ref{eq:f0-perturbate} where a `baseline' density function $f$ is
modulated by a perturbation factor
\begin{equation}
    G(y)=G_0\{h(y)\}   \label{eq:G}
\end{equation}
where $G_0$ is  a univariate distribution function and $h(y)$ is a
real-valued function.
In the discrete case, $f$ will denote a probability function and the
integral in \ref{eq:pi} must be replaced by a summation.

Denote by $Y$ a random variable with density $f$ and by $T$ an independent
variable with distribution function $G_0$.  Assume that a value $y$ sampled
from $f$ is observed conditionally on the event $T \le h(y)$. Then the
observation of a value $y$ generated from $f$ takes place with conditional
probability
\begin{equation}
  \pr{D=1|Y=y} = \pr{T \le h(y)| Y=y}=  G_0\{h(y)\} = G(y)\,,
  \label{eq:Gh}
\end{equation}
while the  unconditional probability of observing a valued from $f$ is
\[
  \pi = \pr{D=1} = \E[Y]{\pr{T \le h(y)| Y=y}} = \E[Y]{ G_0\{h(Y)\}}.
\]
Since in the overwhelming majority of cases the conditional probability
$G(y)$ can reasonably be assumed to be a continuous function of $y$,
continuity is similarly assumed for $G_0$ and consequently for $h$.

It must be underlined that the adoption of the form $G_0\{h(y)\}$ for $G(y)$
in \ref{eq:G} does not constitute a restriction on the latter function,
but only a convenient and often more meaningful way of representing the
conditional probability $G(y)=\pr{D=1|Y=y}$, as seen above for the classical
Heckman formulation. For any arbitrary $G(y)$,
a given choice of $G_0$ identifies a function $h(y)=G_0\inv\{G(y)\}$,
which is unique if $G_0$ is continuous. Clearly, a different choice
of $G_0$ is linked to a different $h$. Which pair $(G_0, h)$ is preferable
for the given $G(y)$ is a component of the modelling process for the problem
at hand; on this step the present proposal allows complete flexibility.

The connection with the literature on symmetry-modulated distributions is
evident both from the expression \ref{eq:f0-perturbate}, which is typical of
that formulation, and from the ensuing stochastic construction via the
independent variables $T$ and $Y$.  There are, however, also some points of
distinction. One is that, as the terms itself suggests, in that literature $f$
typically denotes the symmetric density function of a continuous random
variable, possibly multivariate, and $G_0$ refers to symmetric univariate
random variable; the rare exceptions to this setting appear in recent
non-standard constructions. These symmetry conditions will not be assumed
here.  Another aspect, although of lesser conceptual importance, is the
requirement that $h(y)$ is an odd function with respect to point of symmetry
of $f$ and $G_0$.  Combined with the earlier assumptions, this condition
ensures that the normalizing factor \ref{eq:pi} is $1/2$, with a
major analytical simplification.
This condition is not universal; for instance, it does
not hold for the extended skew-normal distribution in \ref{eq:ESN-pdf} and
\ref{eq:ESN-pdf-new}.  However, it applies to a large fraction of the
literature of symmetry-modulated distributions, but it would be unrealistic
in the present context.

An expression of type \ref{eq:f0-perturbate} can be viewed as the product
of a `baseline' density $f(\cdot)$, which represents the sampling
distribution before censoring takes place, modulated by a perturbation
factor  $G(y)=G_0\{h(y)\}$ which represents the conditional probability
of observing a value $y$ generated by $f(\cdot)$.
From the qualitative viewpoint, adoption of the formulation based on
expression \ref{eq:f0-perturbate} has the advantage of separating, both
conceptually and operationally, the choice of the uncensored distribution
$f$ and the one of the selection mechanism, expressed by the function
$G$. Any choice of $f$ can be combined with any choice
of $G$.

If $y_i$ and $d_i$ denote the analogous quantities of those appearing in
\ref{eq:logL-normal} with an obvious adaptation to the current formulation,
in particular taking into consideration \ref{eq:Gh},
the log-likelihood function takes the form
\begin{eqnarray}
 \log L &=& \sum_{d_i=1}
            \log\left[\pr{D_i=1} \times  f(y_i|D_i=1)\right] +
            \sum_{d_i=0}\log\:\pr{D_i=0}            \label{eq:logL-raw}\\
        &=& \sum_{d_i=1} \log\left[f(y_i) \times \pr{D_i=1|y_i}\right] +
            \sum_{d_i=0}\log\:\pr{D_i=0}                     \nonumber \\
        &=& \sum_{d_i=1} \log\{f(y_i)\:G(y_i)\}  +
            \sum_{d_i=0}\log\left(1- \pi_i\right)       \label{eq:logL}
\end{eqnarray}
where $\pi_i$ denotes the value of \ref{eq:pi} evaluated for the $i$th
individual and a similar dependence on the index $i$ holds for other
components, although not explicit in the notation, as remarked in
connection with \ref{eq:ESN-pdf}.
Correspondingly, $\log L$ depends on parameters which appear in
the  ingredients $f$ and $G$.
As it is typical in similar cases, optimization of \ref{eq:logL} to
obtain maximum likelihood estimates (MLE) must be performed by
numerical methods.

In the development below, we shall examine some specific constructions within
the above scheme, where the ingredients $f$, $G_0$, $h$ are chosen with the
aim of retaining a reasonable algebraic and numerical tractability.
Hopefully, this simplicity should facilitate a meaningful interpretation
from  the applied viewpoint.  There is no attempt, however, to present a
systematic survey of the vast set of all the possible options.

\subsection{Binary response variables} \label{s:binary}

For expository convenience, it seems best to start from the conceptually
simple case of a binary response, yet an important situation from the
applied viewpoint.

Conventionally, the success and failure (uncensored) outcome on the $i$th
subject are associated to a random variable, $Y_{i}$, taking on
values 1 and 0, respectively.
Typically, the probability of success  is expressed as a function
of covariates $x_i$ via a form like
\begin{equation}
  \mu_i =  \E{Y_{i}}=\pr{Y_{i}=1} = P_0(x_i\T \beta)
  \label{eq:mu-binary}
\end{equation}
where $P_0$ is some distribution function on the real line.
The more common options are the logistic and the normal distribution
function, namely
\begin{equation}
  P_0(u) = \frac{\exp(u)}{1+ \exp(u)}
  \qquad\hbox{and}\qquad
  P_0(u) = \Phi(u)\,,
  \label{eq:P0}
\end{equation}
leading to the logit and the probit model for $\mu_i$, respectively.
Alternative choices for $P_0$ are
discussed in the literature on generalized linear models (GLMs).
The probability function of $Y_i$ is then
\begin{equation}
   f(y) = (1-\mu_i)^{1-y}\:\mu_i^y, \quad\qquad y=0, 1\,.
   \label{eq:f-binary}
\end{equation}

One route for modelling selective sampling is via the introduction of a
bivariate normal distribution, similar to \ref{eq:(Y,U)}, followed by
dichotomization of its components, leading to two correlated probit models.
As mentioned earlier, this is the logic followed by Van de Ven and Van Praag
(1981). To derive an inferential technique, the initial part of their
exposition develops an approximate correction factor similar to the one of
Heckman for normal variates, but their subsequent equation (19) presents the
exact likelihood expression, which can be recognized to be analogous to our
\ref{eq:logL-raw}.  In particular, the bivariate normal integrals appearing
in their (19) match the joint probabilities inside the square brackets in
our \ref{eq:logL-raw}.

In this  log-likelihood function, we can convert
the joint probabilities in the first summation of \ref{eq:logL-raw} into
equivalent expressions like those in \ref{eq:logL}.
The term $G(y_i)$ can be expressed via the distribution function of an
extended skew-normal distribution,  similar to the one in \ref{eq:ESN-pdf}
but with reversed role of the underlying continuous variables;
an expression of the required distribution function is given in Section 2.2.3
of Azzalini and Capitanio (2014).

The resulting expression for $G(y_i)$ would be, however, quite involved.
A simpler route is to write directly a model for $G(y_i)$, moving away
from the assumption of an underlying bivariate normal variable. This means
that we regard $Y_i$ as a binary random variable with probability function
\ref{eq:f-binary}, where $\mu_i$ is as in \ref{eq:mu-binary}, and we introduce
suitable ingredients $T\sim G_0$ and $h(\cdot)$ to express the conditional
probability \ref{eq:Gh} of observation. In all cases, computation of $\pi_i$
is elementary for binary response variables; specifically, \ref{eq:pi} becomes
\[
  \pi_i = (1-\mu_i)\:G(0) + \mu_i\:G(1) \,.
\]

In an ideal situation where subject-matter considerations in a given applied
problem indicate an appropriate formulation for $G(y)$, this route
should be followed. Here we discuss some general-purpose options, driven more
by considerations of simplicity, rather than linked to a particular applied
problem.

A necessary requirement for $h(\cdot)$ is to incorporate the covariates
$w_i$ and the simplest way of expressing this is via $\tau_i$, defined
in \ref{eq:(Y,U)}. Formulations that arise naturally for consideration are
a linear expression for $h(y)$ and $T\sim \N(0,1)$, leading to
expressions such as
\begin{equation}
  G(y) = \Phi(\tau_i + \alpha\,y) \quad \hbox{or} \quad
  G(y) =  \Phi(\tau_i + \alpha\mu_i\inv\,y)
  \label{eq:G=Phi(linear)}
\end{equation}
where $\alpha\in\Real$ is a parameter which regulates the dependence on $y$
and the second form introduces a form of standardization, in the sense that
$\E{\mu_i\inv Y_i}=1$; we shall denote $\eta_i=\alpha/\mu_i$.

However, in the present context, there is no compelling reason to stick
to the assumption of normality; this is, in fact,  often made for
reasons like mathematical convenience or widespread familiarity rather
than real belief.  A mathematically simple
alternative is to assume that $T$ has a logistic distribution;
this amounts to replace $\Phi$ in \ref{eq:G=Phi(linear)} by $P_0$
given  in the first expression in \ref{eq:P0}.
Another simple option is to say that $T$ has  an exponential variable
with some fixed parameter,  such as $\E{T}=1$;  we then write
$T\sim\Expn(1)$. In this case,
to ensure that its distribution function is evaluated at positive
values of the argument, we exponentiate the earlier expression
of $h$, arriving at
\begin{equation}
  G(y) = 1-\exp\{-\exp(\tau_i + \alpha\,y)\} \quad \hbox{or} \quad
  G(y) = 1-\exp\{-\exp(\tau_i + \eta_i\,y)\}\,,
  \label{eq:G=Gumbel(linear)}
\end{equation}
which are related to the Gumbel distribution function.

Whatever the adopted form for $G(t)$, an ingredient of interest is
a measure of association between $Y$ and $D$. For a  $2\times2$
probability table such as
\[
  q_{rs} =\pr{Y=r, D=s}, \qquad\quad r=0,1,\quad s=0,1.
\]
a classical measure of dependence is given by the log-odds ratio
\[
  \lambda = \log\frac{q_{00} \:q_{11}}{q_{10}\:q_{01}} \,.
\]
A simple computation lends
\[  q_{00} = \pr{Y=0, D=0} = \pr{Y=0} \: \pr{T\ge h(0)}
           = \pr{Y=0} \: \{1-G(0)\}
\]
and from similar computations one obtains the other probabilities,
arriving at
\[
  \lambda = \log\frac{[1-G(0)]\:G(1)}{G(0)\:[1-G(1)]},
\]
which, recall, depends on the index $i$.

\subsection{Other distributions for the response variable}
\label{s:other-distr}

Among other types of data arising in applications, an important case occurs
when the response variable $Y$ represents count data.  The simpler form of
treatment is via the assumption of a Poisson distribution; for the $i$th
subject we then write
\[
   Y_i\sim\Poisson(\mu_i)
\]
where $\mu_i$ denotes the mean value.
The commonly used form of function relating the mean value to the
covariates is
\begin{equation}
  \mu_i = \log(x_i\T \beta)
  \label{eq:mu-log}
\end{equation}
but also in this case others choices are possible.

As for the selection mechanism, we can still consider those introduced
for binary data, such as \ref{eq:G=Phi(linear)} or some others mentioned
in the subsequent paragraph.

The normalizing constant \ref{eq:pi} is now represented by an infinite sum.
This can be  approximated by a truncated sum:
\[
   \sum_{k=0}^{K} \frac{e^{-\mu_i}\:\mu_i^k}{k!}\: G(k)\,,
\]
where the truncation point $K$ is somewhat larger than the maximal value of
$y_i$. A variant option is to fix a common value $K$ across the whole set of
the $y_i$'s.

The scheme considered so far for the binary and the Poisson distribution can
be employed with some other distribution of the response variable. For
instance, in cases where the Poisson distribution does not provide an adequate
description of the data behaviour, a common solution is to replace it by a
Negative Binomial distribution whose mean value can again be expressed as in
\ref{eq:mu-log} and an additional parameter regulates dispersion.  For our
construction, hardly anything is changed in this switch.

Another situation not feasible for the Gaussian assumption is represented by
positive continuous response variables.  Similarly to the framework of
generalized linear models, it is then quite natural to adopt a distributional
assumption such as the Exponential, the Gamma and the Inverse Gaussian family;
however, this list does not intend to rule out other possibilities.  Again,
the modelling of the selection mechanism can be formulated via one of the
expressions for $G(\cdot)$ which we have examined above.  In these cases, an
operational issue is whether the integral in \ref{eq:pi} allows an explicit
expression. If this is not feasible, as typically it will be the case, we can
still proceed via numerical integration, at the cost of an higher
computational burden.

\subsection{Other forms of selection mechanism} \label{s:other-select}

In the earlier sections, we have discussed various choices of $G(t)$ for
expressing the selection mechanism. These are by no means the only ones,
however. In the case of a non-negative response variable $Y$, an interesting
alternative is provided by the distributional assumption that $T\sim\Expn(1)$
combined with the linear form
\[
  h(y)= \exp(\tau) + \alpha\mu\inv\,y = \lambda + \eta\,y \,,
\]
say, leading to
\begin{equation}
   G(y) = 1- \exp\{- (\lambda + \eta\,y)\}\,.
   \label{eq:G-expn}
\end{equation}

A limitation of this choice is that we need to introduce the condition
$\alpha\ge0$ to ensure that the argument of \ref{eq:G-expn} is positive.
However, if such an assumption on $\alpha$ is plausable on the basis of
subject-matter considerations, then it offers the advantage of an explicit
expression for \ref{eq:pi}, in the wide range of cases where we have
available a similarly explicit expression for the moment generating
function of $Y$; denote it by $M(\cdot)$. It is then immediate to write
\begin{equation}
   \pi = \int_0^\infty f(y) \left(1- e^{-\lambda -\eta y}\right) \d{y}
       =  1- e^{-\lambda} \:M(-\eta)
   \label{eq:pi-explicit}
\end{equation}
where, as usual, in the discrete case the integral sign must be interpreted
as a summation.

For the distributions of $Y$ examined above, that is, binary and Poisson,
use of \ref{eq:pi-explicit} lends
\[
  \pi = 1 -  e^{-\lambda}\: \left[1+\mu(e^{-\eta}-1)\right]
  \qquad\hbox{and}\qquad
  \pi = 1 - \exp\left[-\lambda +\mu\,(e^{-\eta}-1)\right]
\]
but there are many other distributions for which $M(\cdot)$ is known in
closed form, such as the Negative Binomial, Gamma, Inverse Gaussian,
Binomial with arbitrary number of replicates and others more.

There are two reasons why our exposition has not focused on the form
\ref{eq:G-expn}. One is the already-mentioned restriction that $\alpha\ge0$,
which prevents it from general usage. The other reason is that some
numerical exploration has shown that the log-likelihood function \ref{eq:logL}
has, in some cases, an unpleasant behaviour.  For instance,
$\log~L$ can be monotonic, with a maximum at $\alpha=0$ or at
$\alpha\to\infty$.  However, while not appropriate for general usage,
the form \ref{eq:G-expn} may be suitable for specific situations.

\subsection{Computational and additional inferential aspects}

For the numerical maximization of the log-likelihood function, we have
employed the profile log-likelihood function for $\alpha$, namely
\[
  \log L_p(\alpha) = \log L(\alpha, \hat\theta(\alpha))\,,
\]
where $\theta=(\beta\T,\gamma\T)\T$ combines the
two sets of  parameters and $\hat\theta(\alpha)$
is the choice of $\theta$ which maximizes $\log L$ for a given value
of $\alpha$.  The point $\hat\alpha$ which maximizes $\log L_p(\alpha)$
and the corresponding vector $\hat\theta=\hat\theta(\hat\alpha)$
represent the MLE. In the graphical displays below,
we follow the common practice of considering the so-called relative
version of the log-likelihood, which in practice amounts to shift
vertically $\log L_p(\alpha)$ so that its maximum value is~$0$.

To obtain initial values for the numerical search of $\theta$, we fix
initially $\alpha=0$, which amounts to consider two separate generalized
regression models for $Y$ and $D$, free from the sample selection
problem. This produces  estimates of $\beta$ and $\gamma$ to start the
subsequent overall optimization.

For any given $\alpha$, the vector $\hat\theta(\alpha)$ is  obtained
by a separate numerical optimization. This can lead to a substantial
computational burden if a fine grid of $\alpha$ values is scanned.
Usually, a substantial  improvement  in the efficiency of the numerical
search is obtained if an explicit expression of the gradient
\begin{equation}
  \frac{\d}{\d\theta} \log L(\alpha, \theta)
  \label{eq:score}
\end{equation}
is supplied to the optimization algorithm.  General algebraic expressions for
computing first and second order derivatives of the log-likelihood are given
in the appendix. These need to be suitably specified for the adopted choice of
$f$, $G_0$ and $h$.

By standard asymptotic theory, a confidence set for $\alpha$
can be obtained as the set of values satisfying
\begin{equation}
  2\;\left[ \log L_p(\hat\alpha) - \log L_p(\alpha)\right] \le q
  \label{eq:alpha-interval}
\end{equation}
where $q$ denotes the  quantile of the $\chi^2_1$ distribution
function at the chosen confidence level.

Standard errors for $\hat\theta$ can be obtained from
the second-order derivatives matrix evaluated at
$\hat\alpha$, namely
\begin{equation}
   -\left.\frac{\d^2}{\d \theta\d\theta\T} \:
   \log L(\hat\alpha, \theta)\right|_{\theta=\hat\theta} \,.
   \label{eq:obs-info}
\end{equation}
When the score function is not available in an explicit form,
this matrix can be obtained by numerical second order
differentiation of $\log L(\hat\alpha, \theta)$ at $\hat\theta$.
Since expression \ref{eq:obs-info} treats $\alpha$ as fixed at
$\hat\alpha$, it does not fully reflect the variability
involved in the estimation process. However, this limitation
affects only the one-dimensional parameter $\alpha$ and can
reasonably assumed to be of minor importance for the assessment
of standard errors of $\hat\theta$.

\section{Numerical illustrations}

\subsection{German doctor visits} \label{s:doctor}

To illustrate the practical working of the proposed formulation, we make use
of some classical datasets, repeatedly used in the specialized literature.
For the case of binary response variable, we consider data presented by
Riphahn, Wambach and Million (2003) from a longitudinal study concerning user
preferences and usage of the German health insurance system.

We use a subset of these data to parallel the analysis presented in
Example~19.13 of Greene (2012) for the binary response variable $Y$
`defined to equal 1 if an individual makes at least one visit to the
doctor in the survey year', taking into account another binary
variable which indicates whether the individual has subscribed a
``public'' health insurance. For a certain selection of covariates,
the bivariate probit model of Van de Ven \&  Van Praag (1981)
has been fitted to the data and the outcome is presented in Table~19.9
of Greene (2012).

We follow largely the same route, with some differences.
One is to use the logit instead of the probit model for $Y$,
but this is known to have little numerical effect.
For the sample selection mechanism, we obviously considered the one
described above. Specifically, we considered two variant forms,
defining $G$ as follows:
(A) the second expression of \ref{eq:G=Phi(linear)},
(B) the second expression of \ref{eq:G=Gumbel(linear)}.
Another difference is that, taking into account the longitudinal nature
of the study, we only considered the first year of observation for each
subject, to avoid the treatment of multiple observations taken
on the same subject

Our numerical findings are summarized in Table~\ref{t:doctor} and
the two variants of profile log-likelihood function are displayed in
Figure~\ref{t:doctor}. The most noticeable feature is the close similarity
between the outcomes of the two variant forms, both in the numerical
and in the graphical exhibit. Specifically, in case A, we obtained
$\hat\alpha=-2.93$ with a 95\%-level confidence interval $(-4.92, -1.70)$
using \ref{eq:alpha-interval};
in case B,  $\hat\alpha=-3.07$ with confidence interval $(-5.40, -1.70)$.
Also the values of $\hat\theta$ and their standard errors reported in
Table~\ref{t:doctor} are very similar in the two cases.

The closeness of the two sets of results is reassuring, especially
in the light of the recurrent criticism of Heckman formulation for
its instability with respect to the assumption on the underlying
stochastic ingredients.
If one has to choose between the two models, variant A has maximized
log-likelihood $-6510.03$ versus $-6514.43$ for variant $B$; hence
A would be preferable according to Akaike and similar information
criteria.

The values in Table~\ref{t:doctor}  are also broadly similar to those
in Table~19.9 of Greene (2012). The largest differences occurs in the
two intercept terms, but these are not important for interpretation;
the other terms give  fairly similar indications although with some
differences.

\begin{figure}
\caption{\sl German doctor visits data with logit model for the response
variable and two choices of the  selection mechanism:
(A) $T\sim\N(0,1)$, $h(y)=\tau+\eta y$,
(B) $T\sim\Expn(1)$, $h(y)=\exp(\tau+\eta y)$.}
\label{f:doctor}
\centerline{
   \includegraphics[width=0.49\hsize]{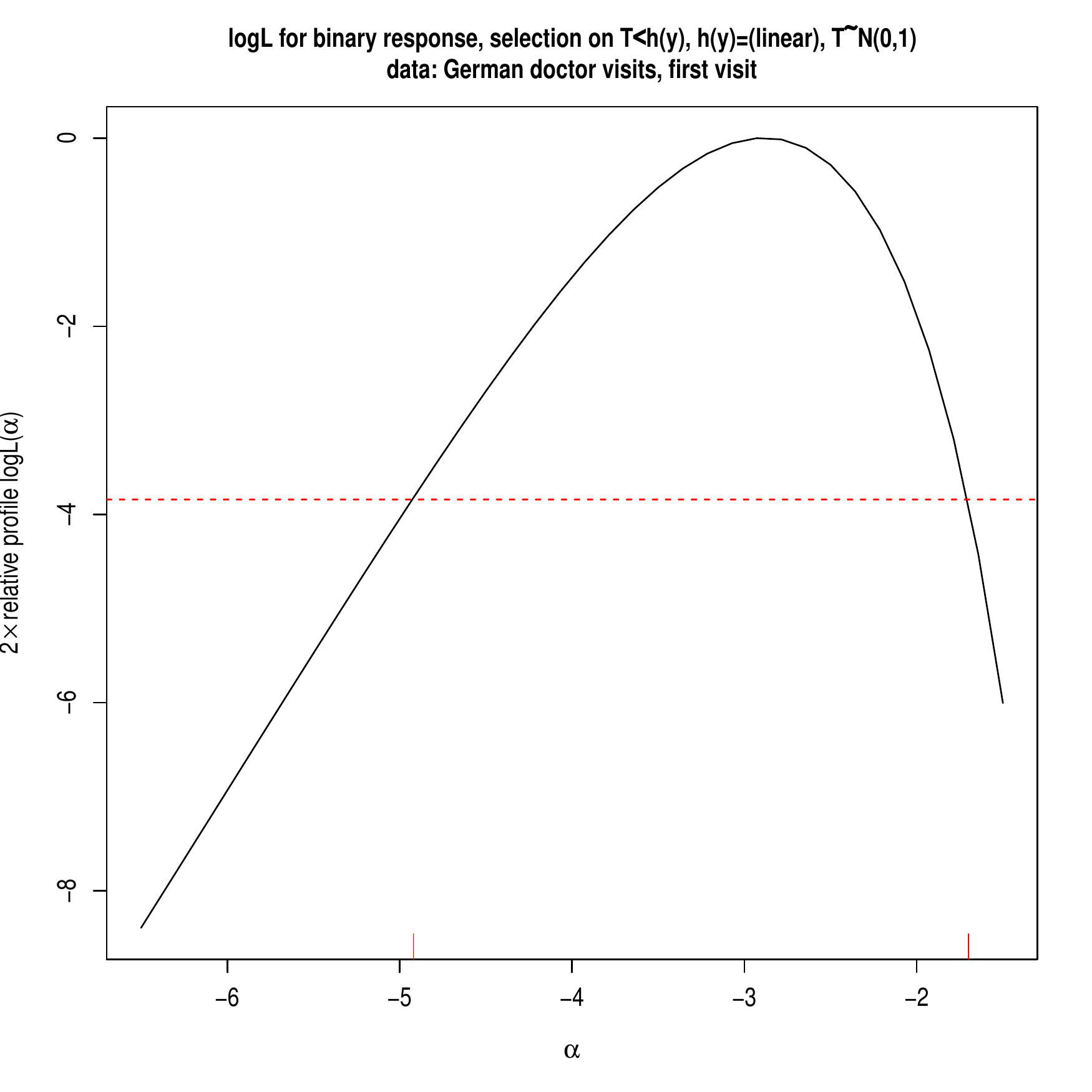}\quad
   \includegraphics[width=0.49\hsize]{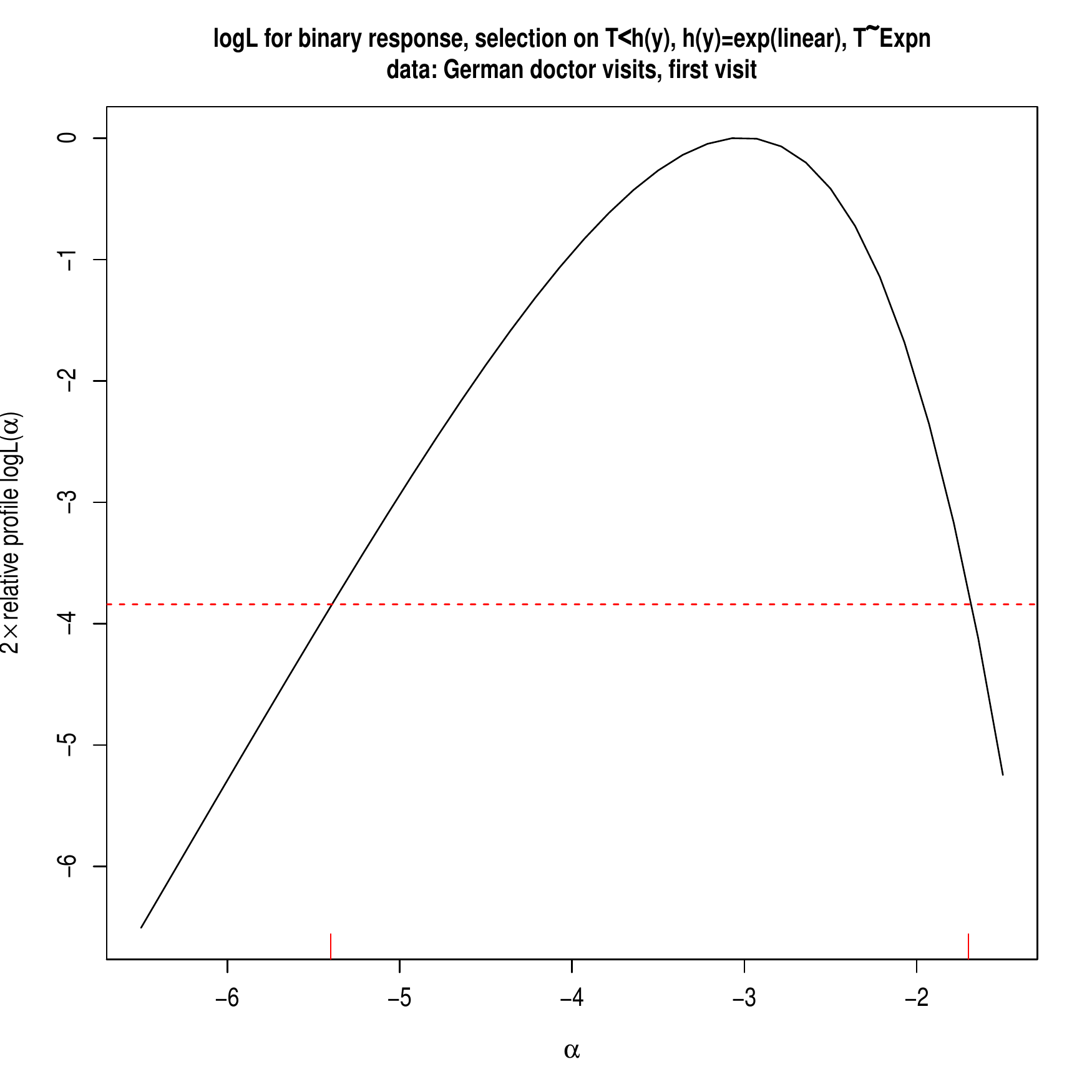}}
\end{figure}
\begin{table}
\centering
\caption{\sl German doctor visits data with logit model for the response
variable and two choices of  the  selection mechanism:
(A) $T\sim\N(0,1)$, $h(y)=\tau+\eta y$,
(B) $T\sim\Expn(1)$, $h(y)=\exp(\tau+\eta y)$.}
\label{t:doctor}
\par\vspace{3ex}
(A)  maximized $\log L=-6510.03$,  $\hat\alpha= -2.93$
with 95\%-level confidence interval $(-4.92, -1.70)$
\par\vspace{1ex}
logit model for the response variable
\begin{tabular}{rrrrrrr}
  \hline
 & one & age & income & kids & education & married \\
  \hline
$\hat\beta$ & -0.49 & 0.0158 & -0.31 & -0.149 & 0.059 & -0.045 \\
  \texttt{std.err} &  0.15 & 0.0019 &  0.05 &  0.029 & 0.010 &  0.032 \\
  \texttt{ratio} & -3.28 & 8.2750 & -5.79 & -5.208 & 5.707 & -1.383 \\
   \hline
\end{tabular}
\par\vspace{1ex}
selection model\par
\begin{tabular}{rrrrr}
  \hline
 & one & age & education & female \\
  \hline
$\hat\gamma$& 9.54 & -0.024 & -0.276 & 0.29 \\
  \texttt{std.err}& 0.26 & 0.003 & 0.016 & 0.05 \\
  \texttt{ratio} & 36.87 & -7.205 & -17.160 & 6.18 \\
   \hline
\end{tabular}
\par\vspace{3ex}
(B)  maximized $\log L=-6514.43$,  $\hat\alpha=-3.07$
with 95\%-level confidence interval $(-5.40, -1.70)$
\par\vspace{1ex}
logit model for the response variable
\begin{tabular}{rrrrrrr}
  \hline
 & one & age & income & kids & education & married \\
  \hline
$\hat\beta$ & -0.57 & 0.0157 & -0.31 & -0.109 & 0.064 & -0.045 \\
  \texttt{std.err} & 0.15 & 0.0018 & 0.05 & 0.023 & 0.010 & 0.026 \\
  \texttt{ratio} & -3.81 & 8.6514 & -5.63 & -4.689 & 6.118 & -1.715 \\
   \hline
\end{tabular}
\par\vspace{1ex}
selection model\par
\begin{tabular}{rrrrr}
  \hline
 & one & age & education & female \\
  \hline
$\hat\gamma$ & 9.46 & -0.026 & -0.272 & 0.23 \\
  \texttt{std.err} & 0.28 & 0.003 & 0.018 & 0.04 \\
  \texttt{ratio} & 34.19 & -8.031 & -15.403 & 5.76 \\
   \hline
\end{tabular}
\end{table}

\subsection{Credit cards derogatory reports}

Greene (1998) examines a number of aspects in automatic credit-scoring
methodology to scrutinize applications for financial credit in order to
discard those which are particularly exposed to the risk of default or some
other critical behaviours.  In a context where a large number of such
applications arise in a given time period, the adoption of an automated system
is required for such scrutiny.  A good example of this situation is provided
by applications for credit cards, which are typically evaluated in an
automated way on the basis of historical data.  As the author notes, `In order
to enter the sample used to build the model, an individual must have already
been `accepted'' (p.\,299) with the implication that `a predictor of default
risk in a given population of applicants can be systematically biased because
it is constructed from a nonrandom sample of past applicants, that is, those
whose applications were accepted.' (p.\,300).  Consequently, he advocates to
take into consideration the sample selection mechanism, by including into
consideration also subjects whose application had not been approved.

The opening sentence of Greene (1998, Section 5) is: `By far the most
significant variable on the card-holder equation is MDRs, the number of major
derogatory reports'; this is the response variable $Y$ considered below.
Greene's treatment of the problem was based on a formulation similar to the
one of Terza (1998), mentioned in Section~\ref{s:non-gaussian} above, which
involves the introduction of an extra latent variable $\eps$. Another issue is
that some of the covariates employed in this formulation are not included in
the dataset available to us. Therefore a direct comparison with our treatment
described next is not possible.

\begin{figure}
\caption{\sl Credit cards derogatory reports with log-linear model
for the mean value of the Poisson response
variable and two choices of the  selection mechanism:
(A) $T\sim\N(0,1)$, $h(y)=\tau+\eta y$,
(B) $T\sim\Expn(1)$, $h(y)=\exp(\tau+\eta y)$.}
\label{f:creditcards}
\centerline{
   \includegraphics[width=0.49\hsize]{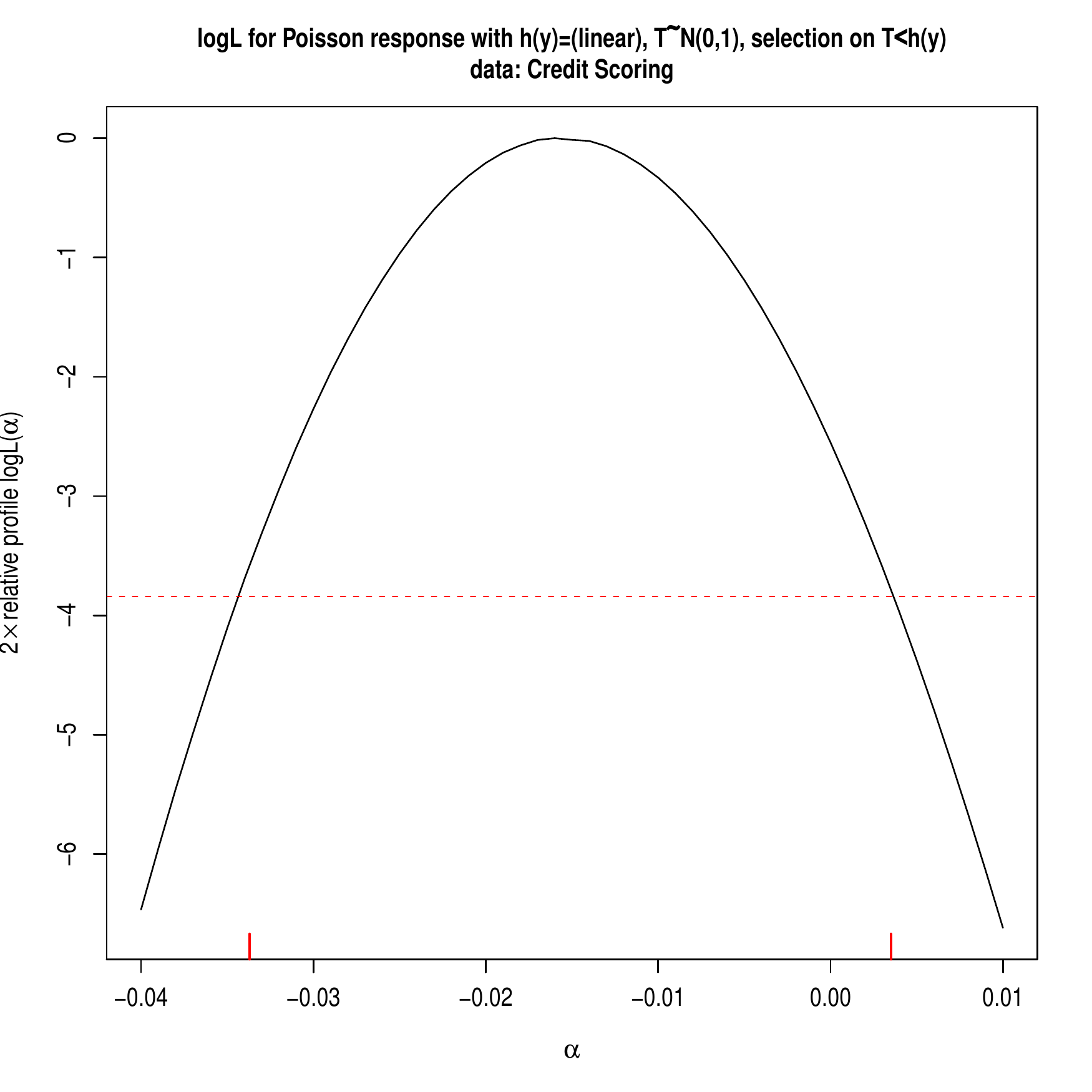}\quad
   \includegraphics[width=0.49\hsize]{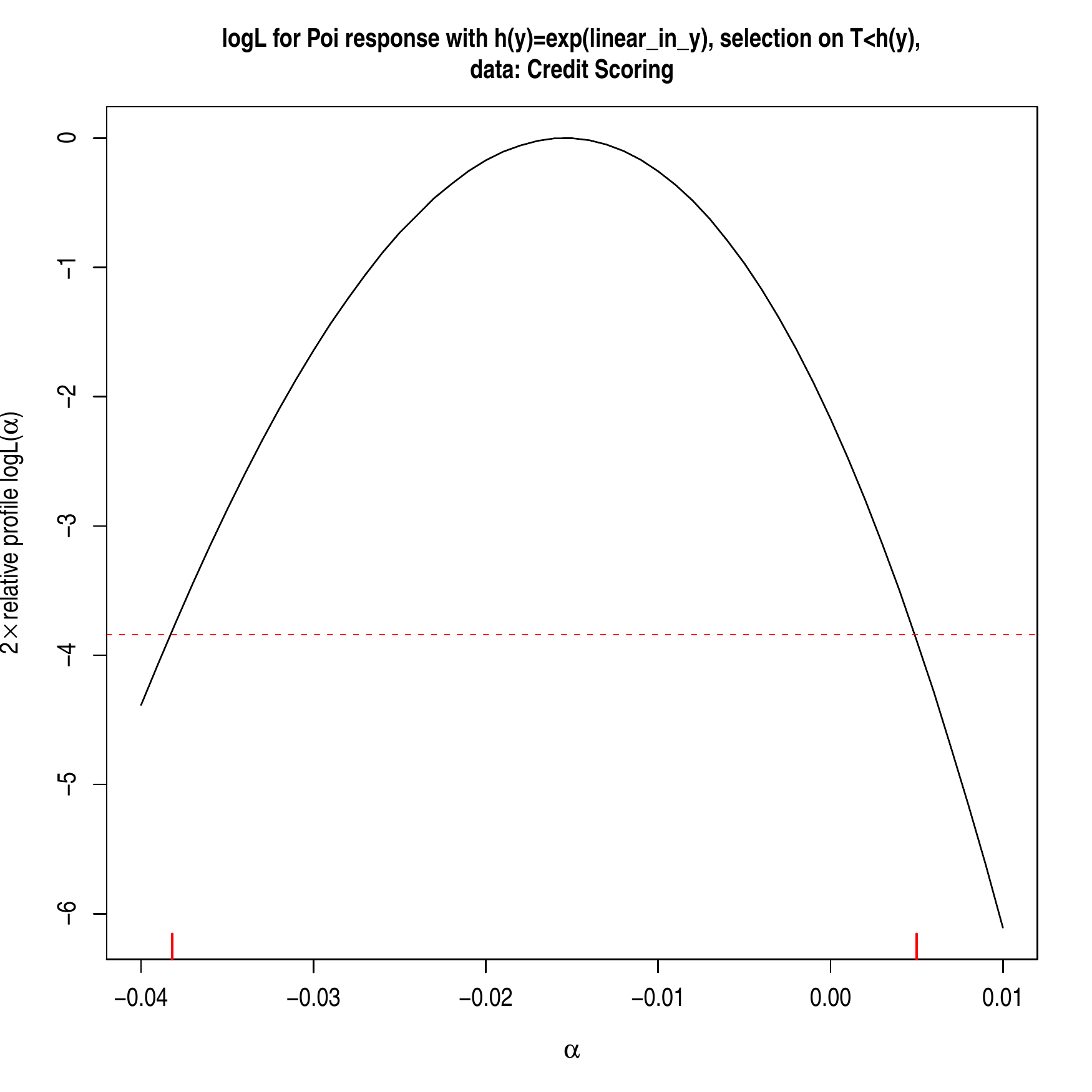}}
\end{figure}

For our formulation, two choices of the selection mechanism have been
considered for these data, namely the same employed in Section~\ref{s:doctor}.
Figure~\ref{f:creditcards} and Table~\ref{t:creditcards} provide the
summary outcome of the numerical work, in the form of profile log-likelihood
function, MLEs and standard errors. Also in this example the log-likelihood
has a smooth nearly-quadratic behaviour for both variants of the selection
model.  Again, MLEs and their standard errors are in close agreement in the
two variants, A and B.

\begin{table}

\centering
\caption{\sl Credit cards derogatory reports with log-linear model
for the mean value  of the Poisson response
variable and two choices of the  selection mechanism:
(A) $T\sim\N(0,1)$, $h(y)=\tau+\eta y$,
(B) $T\sim\Expn(1)$, $h(y)=\exp(\tau+\eta y)$.}
\label{t:creditcards}
\par\vspace{3ex}
(A) maximized $\log L=-11387.63$,  $\hat\alpha= -0.016$
with  95\% confidence interval $(-0.0337, 0.0035)$
\par\vspace{1ex}
log-linear model for the response variable\par
\begin{tabular}{rrrrr}
  \hline
 & const & Age & Income & Exp\_Inc  \\
  \hline
$\hat\beta$      & -3.22  &  0.0210 &  0.165 & 1.23 \\
\texttt{std.err} &  0.09  &  0.0023 &  0.016 & 0.16 \\
\texttt{ratio}   & -35.90 &  9.1867 & 10.294 & 7.80 \\
       \hline
\end{tabular}
\par\vspace{1ex}
 selection model\par
\begin{tabular}{rrrrrrr}
  \hline
    &   Const & Age & Income & Ownrent & Adepcnt & Selfempl\\
  \hline
  $\hat\gamma$     & 0.36 & -0.0014 &  0.217 & 0.224 & -0.114 & -0.343 \\
  \texttt{std.err} & 0.05 &  0.0013 &  0.013 & 0.028 &  0.010 &  0.051 \\
  \texttt{ratio}   & 7.87 & -1.0308 & 17.330 & 8.044 & -11.100 & -6.681 \\
   \hline
\end{tabular}
\par\vskip 3ex
(B)
maximized $\log L= -11399.83$, $\hat\alpha= -0.015$
with 95\% confidence interval $(-0.0382, 0.005)$
\par\vspace{1ex}
log-linear model of response variable\par
\begin{tabular}{rrrrr}
  \hline
       & const &  Age & Income & Exp\_Inc  \\
  \hline
  $\hat\beta$       & -3.22 & 0.0209 &  0.166 & 1.23 \\
   \texttt{std.err} &  0.09 & 0.0023 &  0.016 & 0.16 \\
  \texttt{ratio}   & -35.88 & 9.1672 & 10.360 & 7.78 \\

   \hline
\end{tabular}
\par\vspace{1ex}
selection model\par
\begin{tabular}{rrrrrrr}
  \hline
    & const &  Age & Income & Ownrent & Adepcnt & Selfempl \\
  \hline
  $\hat\gamma$     & 0.09 & -0.0012 &  0.170 & 0.204 &  -0.098 & -0.32 \\
  \texttt{std.err} & 0.04 &  0.0012 &  0.010 & 0.024 &   0.009 &  0.05 \\
  \texttt{ratio}   & 2.29 & -1.0542 & 17.677 & 8.357 & -10.415 & -6.67 \\
   \hline
\end{tabular}
\end{table}

\section{Concluding remarks}

The proposed formulation encompasses a wide range of choices for the
distribution of response variable and for the sample selection mechanism.  A
feature which seems appealing to us is the complete separation of these two
ingredients, which can be chosen independently from each other, unlike some
existing proposals. Another aspect of conceptual simplicity is that our
formulation involves only one latent variable in the selection mechanism,
similarly to the original Heckman proposal.

Our numerical experience has indicated an appealing stability of the
parameters of interest with respect to the choice of the selection mechanism.
Since the range of cases considered here is limited and they are confined
to discrete response variables, this point requires further exploration.
It is conceivable that this stability is a pleasant side-effect of the
discrete nature of the response variable.

We have not fully elaborated on the forms of the linear predictors for
the $\mu_i$ and $\tau_i$, for which we have retained simple parametric
expressions, since our key interest was the developement of the selection
mechanism. It is however possible to introduce more elaborate expressions
such as spline functions, following a line analogous to Marra and Wyszynski
(2016).

\subsection*{Acknowledgments}
Hyoung-Moon Kim's research was supported by Basic Science Research Program through the
National Research Foundation of Korea (NRF) funded by the Ministry of Education
(2015R1D1A1A01059161). Hea-Jung Kim's research was supported by Basic Science Research
Program through the National Research Foundation of Korea (NRF) funded by the Ministry of
Education (2015R1D1A1A01057106).

\section*{References}
\biblioitem
Azzalini, A. with the collaboration of Capitanio, A. (2014).
\emph{The Skew-Normal and Related Families}.
Cambridge University Press, Cambridge.

\biblioitem
Copas, J.B., and Li, H. G. (1997).
Inference for non-random samples (with discussion).
\emph{J. R. Stat. Soc., series B}, \textbf{59}, 55--95.


\biblioitem
Greene, W. (1998).
Sample selection in credit-scoring models.
\emph{Japan and the World Economy} \textbf{10}, 299--316.

\biblioitem
Greene, W. H. (2012). \emph{Econometric Analysis}, 7th edition.
Pearson Education Ltd, Harlow.

\biblioitem
Heckman, J.~J. (1976).
 The common structure of statistical models of truncation, sample selection
 and limited dependent variables, and a simple estimator for such models.
 {\em Ann. Econ. Socl. Measmnt.}, {\bf 5}, 475--492.


\biblioitem
 Heckman, J.~J. (1979).
 Sample selection bias as a specification error.
 \emph{Econometrica}, \textbf{47}, 153--161.

\biblioitem
Marchenko, Y.~V., and Genton, M.~G. (2012).
  A {H}eckman selection-$t$ model.
  \emph{ J.\ Amer.\ Statist.\ Assoc.}, {\bf 107}, 304--317.

\biblioitem
Marra, G. and Wyszynski, K. (2016).
  Semi-paarametric copula sample selection models for count responses.
  \emph{Comp. stat \& Data An.}, \textbf{104}, 110--129.

\biblioitem
McCullagh, P. and Nelder, J. A. (1989).
  \emph{Generalized Linear Models}, 2nd edition.
  Chapman \& Hall/CRC, London.

\biblioitem
Riphahn, R.\,R.,  Wambach, A. and  Million, A. (2003).
Incentive Effects in the Demand for Health Care:
A Bivariate Panel Count Data Estimation.
\emph{Journal of Applied Econometrics}, \textbf{18}, 387--405.

\biblioitem
Terza, J. V. (1998).
Estimating count data models with endogenous switching:
Sample selection and endogenous treatment effects,
\emph{Journal of Econometrics} 84, 129-154.

\biblioitem
Van de Ven, Wynand P.M.M. and  Van Praag, Bernard M.S. (1981).
The demand for deductibles in private health insurance: A probit
model with sample selection.
\emph{Journal of Econometrics}, \textbf{17}(2), p.229--252.
Corrigendum in  Vol.\,\textbf{22}(3), p.\,395 (1983).

\biblioitem
Zhelonkin, M., Genton, G.~G. and Ronchetti, E. (2016).
Robust inference in sample selection models.
\emph{J. R. Stat. Soc., series B}, \textbf{78}, 805--827.

\appendix
\section*{Appendix: score function and Hessian matrix}

For the overwhelming majority of cases of interest in applications, the
density function $f$ is a member of the exponential family which enter the
formulation of generalized linear models; hence we focus on this situation.
Following essentially the notation of McChullagh and Nelder (1989), we write
the baseline density (or probability function, in the discrete case) as
\begin{equation}
   f(y;\vartheta,\psi) = \exp \left\{ \frac{y\vartheta-b(\vartheta)}{a(\psi)}
        + d(y,\psi)\right\}
   \label{eq:f-GLM}
\end{equation}
where $a(\cdot), b(\cdot)$ and $d(\cdot)$ are known functions. In some
cases, the dispersion parameters $\psi$ is known; important instances of this
type are the Poisson and the binomial distribution.
%

On inserting expression \ref{eq:f-GLM} in \ref{eq:logL}, the log-likelihood
function becomes
\begin{equation}\label{eq:LogLikeGLM}
    \log L (\alpha, \theta, \psi) =
       \sum_{d_i=1} \left[ \frac{y_i\vartheta_i-b(\vartheta_i)}{a_i (\psi)}
       + d(y_i,\psi)  + \log G_0\{h(y_i)\}\right]
       + \sum_{d_i=0}  \log(1-\pi_i)
\end{equation}
%
%
whose derivatives with respect to the parameters
$\beta, \gamma, \psi$ are as follows:
\begin{eqnarray*}
s(\beta_j) = \dfrac{\partial\log L (\alpha, \theta, \psi)}{\partial \beta_j}
&=& \sum_{d_i=1} \left[ \frac{y_i-\mu_i}{V_i} + \frac{g_0\{h(y_i)\}}{G_0\{h(y_i)\}} \frac{\partial h(y_i)}{\partial\mu_i}\right] \frac{1}{g'(\mu_i)}x_{ij}   \\
&&- \sum_{d_i=0} \left[ \frac{\partial\pi_i/\partial\mu_i}{1-\pi_i} \right]
\frac{1}{g'(\mu_i)}x_{ij}, \qquad  \hbox{for~} j=1, \cdots, p,
\\
s(\gamma_h) = \dfrac{\partial\log L (\alpha, \theta, \psi)}{\partial \gamma_h}
&=&\sum_{d_i=1} \left[\frac{g_0\{h(y_i)\}}{G_0\{h(y_i)\}} \frac{\partial
    h(y_i)}{\partial\tau_i}\right] w_{ih} - \sum_{d_i=0} \left[
  \frac{\partial\pi_i/\partial \tau_i}{1-\pi_i} \right] w_{ih},
  \qquad  \hbox{for~} h=1, \cdots, q,
\\
s(\psi) = \dfrac{\partial\log L (\alpha, \theta, \psi)}{\partial \psi}
&=& \sum_{d_i=1} \left[\frac{b(\vartheta_i) - y_i\vartheta_i}{a^2_i(\psi)} a'_i(\psi) + \frac{\partial d(y_i, \psi)}{\partial\psi} \right]-  \sum_{d_i=0}
 \frac{\partial\pi_i/\partial\psi}{1-\pi_i}
\end{eqnarray*}
where $V_i=a_i(\psi)b''(\vartheta_i)=\var{Y_i}$,
$\E{Y_i}=\mu_i=b'(\vartheta_i)$,  $g_0=G_0'$ and $g(\mu_i)=x_i\T\beta$ is
called the link function.

%

The second order derivatives of \ref{eq:LogLikeGLM} are given by the following
expressions:
\begin{eqnarray*}
  H(\beta_j, \beta_h) &=&
  \sum_{d_i=1} \left[  - \frac{1}{a(\psi)} +
    \left(\frac{g_0'\{h(y_i)\}}{G_0\{h(y_i)\}} -
      \left(\frac{g_0\{h(y_i)\}}{G_0\{h(y_i)\}} \right)^2 \right) \left(
      \frac{\partial h(y_i)}{\partial \mu_i} \right)^2 b''(\vartheta_i)
  \right.
  \\
  &&  + \frac{g_0\{h(y_i)\}}{G_0\{h(y_i)\}} \left(\frac{\partial^2
     h(y_i)}{\partial \mu_i^2} b''(\vartheta_i) +\frac{\partial
     h(y_i)}{\partial \mu_i} \frac{b'''(\vartheta_i)}{b''(\vartheta_i)} \right)
  \\
  && \left. - \left\{ \frac{y_i -b'(\vartheta_i)}{V_i} + \frac{g_0\{h(y_i)\}}{G_0\{h(y_i)\}} \frac{\partial h(y_i)}{\partial\mu_i} \right\}  \cdot \left\{\frac{b'''(\vartheta_i)}{b''(\vartheta_i)} + \frac{b''(\vartheta_i) g''(\mu_i)}{g'(\mu_i)} \right\} \right] \frac{x_{ih}x_{ij}}{b''(\vartheta_i)(g'(\mu_i))^2}\\
  && + \sum_{d_i=0} \left[ \frac{1}{\pi_i-1} \left(\frac{\partial^2 \pi_i}{\partial \mu_i^2} b''(\vartheta_i) + \frac{\partial \pi_i}{\partial \mu_i} \frac{b'''(\vartheta_i)}{b''(\vartheta_i)} \right) -\frac{1}{(1-\pi_i)^2} \left(\frac{\partial \pi_i}{\partial \mu_i} \right)^2 b''(\vartheta_i) \right. \\
  &&  \left. + \frac{\partial\pi_i/\partial\mu_i}{1-\pi_i} \cdot
    \left\{\frac{b'''(\vartheta_i)}{b''(\vartheta_i)} + \frac{b''(\vartheta_i)
        g''(\mu_i)}{g'(\mu_i)} \right\} \right]
  \frac{x_{ih}x_{ij}}{b''(\vartheta_i)(g'(\mu_i))^2},
  \\
  H(\beta_j, \gamma_h) &=&
  \sum_{d_i=1} \left[ \left\{\frac{g_0' \{h(y_i)\}}{G_0 \{h(y_i)\}} -\left(\frac{g_0 \{h(y_i)\}}{G_0\{h(y_i)\}} \right)^2 \right\}
    \frac{\partial h(y_i)}{\partial \tau_i} \frac{\partial h(y_i)}{\partial \mu_i} \frac{w_{ih} x_{ij}}{ g'(\mu_i)} \right.\\
  && - \left.  \sum_{d_i=0}  \left(\frac{\frac{\partial^2 \pi_i}{\partial \tau_i \partial \mu_i} }{1-\pi_i} +\frac{\partial \pi_i}{\partial \tau_i} \frac{\partial \pi_i}{\partial \mu_i} \frac{1}{(1-\pi_i)^2} \right) \right]
  \frac{w_{ih} x_{ij}}{ g'(\mu_i)},
\\
H(\beta_j, \psi) &=&
\sum_{d_i=1}  \frac{a' (\psi) (\mu_i - y_i)}{a^2_i (\psi) b''
  (\vartheta_i)}\frac{x_{ij}}{g' (\mu_i)}  - \sum_{d_i=0}
\frac{1}{(1-\pi_i)^2} \left\{ \frac{\partial^2 \pi_i}{\partial \psi \partial
    \mu_i} (1-\pi_i) + \frac{\partial \pi_i}{\partial \psi} \frac{\partial
    \pi_i}{\partial \mu_i} \right\}  \frac{x_{ij}}{g' (\mu_i)},
\\
H(\gamma_j, \gamma_h)&=&
\sum_{d_i=1} \left[ \left\{\frac{g_0' \{h(y_i)\}}{G_0 \{h(y_i)\}} -\left(\frac{g_0 \{h(y_i)\}}{G_0\{h(y_i)\}} \right)^2 \right\} \left(\frac{\partial h(y_i)}{\partial \tau_i}\right)^2
 +  \frac{g_0 \{h(y_i)\}}{G_0 \{h(y_i)\}} \frac{\partial^2 h(y_i)}{\partial \tau_i^2} \right] w_{ij} w_{ih}\\
&& -  \sum_{d_i=0} \frac{1}{(1-\pi_i)^2} \left( \frac{\partial^2
    \pi_i}{\partial \tau_i^2} (1-\pi_i) + \left(\frac{\partial \pi_i}{\partial
      \tau_i} \right)^2\right) w_{ij} w_{ih},
\\
H(\gamma_j, \psi) &=&
 -\sum_{d_i=0} \frac{1}{(1-\pi_i)^2} \left\{ \frac{\partial^2 \pi_i}{\partial
     \psi \partial \tau_i} (1-\pi_i) + \frac{\partial \pi_i}{\partial \psi}
   \frac{\partial \pi_i}{\partial \tau_i} \right\} w_{ij},
\\
H(\psi, \psi) &=&
\sum_{d_i=1}  \left\{\frac{2(y_i\vartheta_i -b(\vartheta_i))}{a^3_i(\psi)} (a'_i(\psi))^2 - \frac{y_i\vartheta_i -b(\vartheta_i)}{a^2_i(\psi)}a''_i(\psi) + \frac{\partial^2 d(y_i, \psi)}{\partial \psi^2}\right\}  \\
&& - \sum_{d_i=0} \frac{1}{(1-\pi_i)^2} \left\{ \frac{\partial^2 \pi_i}{\partial \psi^2} (1-\pi_i) + \left( \frac{\partial \pi_i}{\partial \psi} \right)^2 \right\}.
\end{eqnarray*}

\end{document}